\documentclass[aps,pra,showpacs,twocolumn]{revtex4-1}
\usepackage{amsmath}
\usepackage{amscd}
\usepackage{wasysym}
\usepackage[ansinew]{inputenc}
\usepackage[T1]{fontenc}
\usepackage{ae,aecompl}
\usepackage{amsfonts}
\usepackage{amsthm}
\usepackage{dsfont}
\usepackage{graphicx}
\usepackage{amsfonts}

\begin{document}
\title{Breakdown of adiabatic transfer of light in waveguides in the presence of absorption}
\author{Eva-Maria Graefe$^{1}$, Alexei A. Mailybaev$^{2,3}$, and Nimrod Moiseyev$^{4}$}
\address{${}^1$Department of Mathematics, Imperial College London, London, SW7 2AZ, United Kingdom\\
${}^2$Instituto Nacional de Matem\'atica Pura e Aplicada -- IMPA, Estrada Dona Castorina 110, 22460-320 Rio de Janeiro, RJ, Brazil\\
${}^3$Institute of Mechanics, Lomonosov Moscow State University, Russia\\
${}^4$Department of Physics and Minerva Center for Nonlinear Physics of Complex Systems, Technion -- Israel Institute of Technology, Haifa, 32000 Israel}
\begin{abstract}
In atomic physics, adiabatic evolution is often used to achieve a robust and efficient population transfer. Many adiabatic schemes have also been implemented in optical waveguide structures. Recently there has been increasing interests in the influence of decay and absorption, and their engineering applications. Here it is shown that even a small decay can significantly influence the dynamical behaviour of a system, above and beyond a mere change of the overall norm. In particular, a small decay can lead to a breakdown of adiabatic transfer schemes, even when both the spectrum and the eigenfunctions are only sightly modified. This is demonstrated for the generalization of a STIRAP scheme that has recently been implemented in optical waveguide structures. Here the question how an additional absorption in either the initial or the target waveguide influences the transfer property of the scheme is addressed. It is found that the scheme breaks down for small values of the absorption at a relatively sharp threshold, which can be estimated by simple analytical arguments. 
\end{abstract}
\maketitle

\section{Introduction}
The adiabatic theorem of Hermitian quantum mechanics provides the basis for many experimental schemes to manipulate and control quantum systems. Prominent examples from atomic physics include the RAP and STIRAP schemes \cite{Berg98b} for population transfer in effective two or three level systems. In many applications, however, quantum systems have states with finite lifetimes described by complex energies. These more realistic situations cannot be described by conventional Hermitian quantum mechanics, but require an additional anti-Hermitian part in the Hamiltonian (see, e.g., \cite{Mois11,Okol03} and references therein). Due to the analogy of the time dependent Schr\"odinger equation and the paraxial approximation for light propagation in optical media \cite{opt_herm,Longh09}, the dynamics generated by non-Hermitian Hamiltonians can conveniently be realized using optical systems \cite{opt_nherm}. 

It has previously been pointed out in the literature that the adiabatic theorem does not necessarily hold for non-Hermitian systems \cite{Nenc92}, and it has recently been shown that this can lead to new effects in the ``deep'' non-Hermitian regime in the presence of exceptional points \cite{nhermad}. However, it is often assumed that \textit{small} decay rates do not modify the systems behaviour drastically, and their effects on adiabatic behaviour seem hitherto not to have been fully appreciated. 

Here we show that even a small decay rate, which does not modify the static behaviour of a system significantly, can lead to a breakdown of adiabatic transfer properties. We demonstrate this fact for a STIRAP-related scheme, which is readily implemented in optical waveguide structures. In this scheme a sharp threshold for the breakdown of the transfer property is observed, which we calculate to a good approximation using simple analytical arguments. Note that models similar to the one considered here also appear in the context of cavity QED systems \cite{Kuhn}.

\section{The model: A STIRAP-type scheme in three coupled waveguides} 
The STIRAP scheme was originally proposed and implemented for population transfer in three-level atoms \cite{Berg98b,adiab_three}. 
Recently, STIRAP-type schemes have also been implemented in optical waveguides \cite{Lahi08,Dreis09,STIRAP_waveguide1,STIRAP_waveguide2}, where the propagation distance $z$ takes the role of time in conventional quantum systems. The optical realization allows for a straightforward experimental implementation of additional decay of varying strength using absorbing materials. Thus, in what follows we shall proceed our analysis in the waveguide context. The propagation of light in this system is governed by a Schr\"odinger type equation, where the propagation distance $z$ takes over the role of time in conventional quantum systems \cite{opt_herm,Longh09}:
\begin{equation}
\label{eqn_evo}
i\frac{d}{dz}|\psi(z)\rangle=H(z)|\psi(z)\rangle.
\end{equation}
In a system of $N$ waveguides $|\psi(z)\rangle$ is a vector in $\mathds{C}^N$, whose components are the wave amplitudes in the different waveguides, and the Hamiltonian $H(z)$ encodes the refractive index in the waveguides as well as the coupling between them. 

\begin{figure}
\centering
\includegraphics[width=0.25\textwidth]{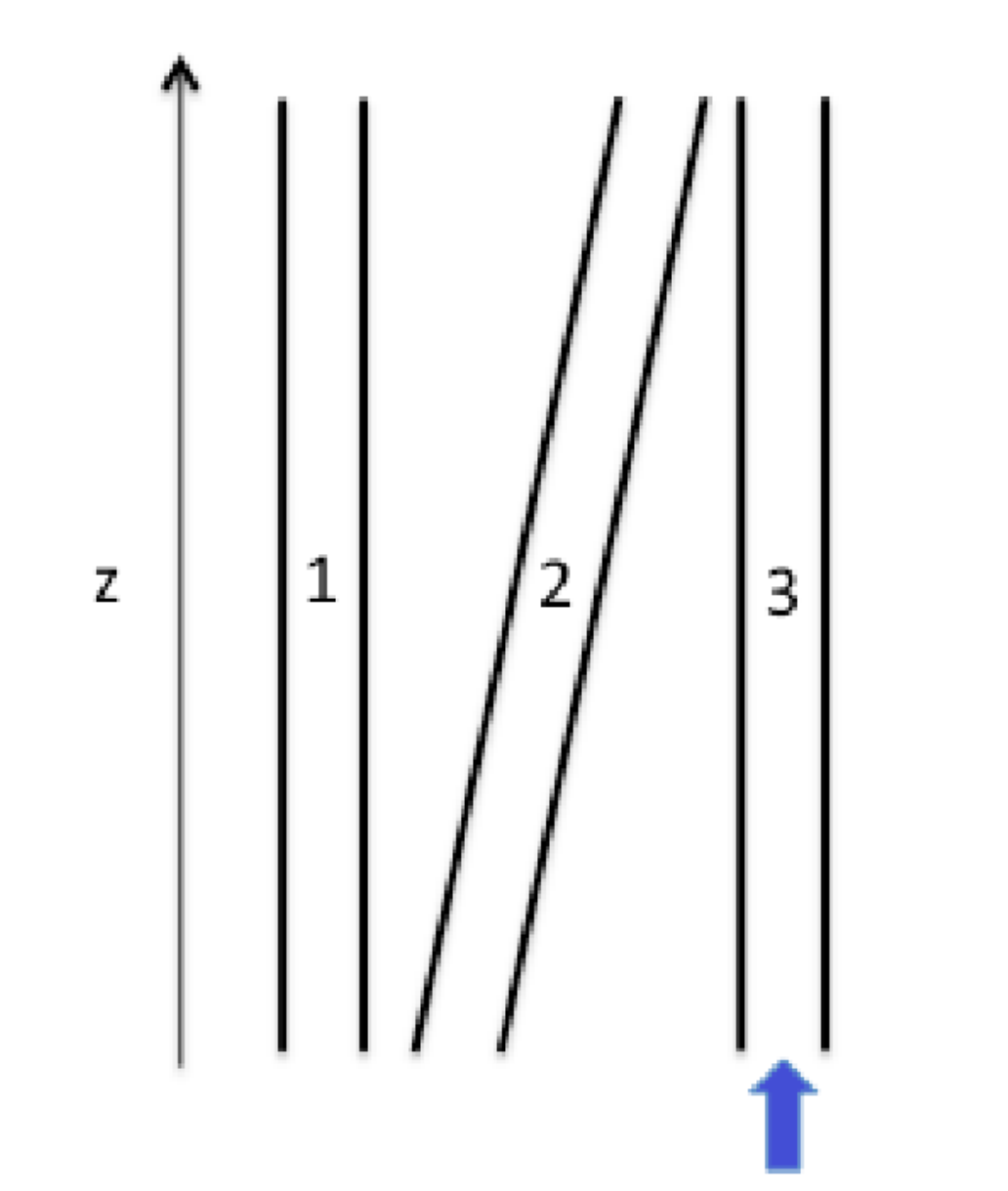}
\caption{Sketch of the STIRAP scheme in three coupled waveguides as implemented in \cite{Lahi08}, leading to adiabatic transfer of light from waveguide 3 to waveguide 1.}
\label{fig0}
\end{figure}

In the following, we study a generalization of the setup investigated in \cite{Lahi08}, which consists of two parallel (left and right) waveguides with an additional diagonally directed central waveguide, as schematically depicted in Fig. \ref{fig0}.
Since the couplings between neighbouring waveguides depend exponentially on their distances, this system can be described by a Hamiltonian of the type 
\begin{equation}
H(z) = \left(\begin{array}{ccc}
0 & \,v(z) & 0 \\
\,\,\,v(z) & 0 & w(z) \\
\,\,\,0 & \,w(z) & 0
\end{array}\right),
\label{eq_Ham}
\end{equation} 
with couplings of the form
\begin{equation}
v(z) = 1/w(z) = {\rm e}^{-a(z-L/2)/L}.
\label{eq2}
\end{equation}
The eigenvalues of the system for a given value of $z$ are given by
\begin{equation}
E_0 = 0, \quad E_{\pm} = \pm \omega, \quad {\rm with}\ \omega=\sqrt{v^2+w^2},
\label{eq4}
\end{equation}
with the corresponding eigenstates
\begin{equation}
|\varphi^0\rangle = \left(\begin{array}{c} \cos\theta \\ 0 \\ -\sin\theta \end{array}\right),\quad
|\varphi^\pm\rangle = \frac{1}{\sqrt{2}}\left(\begin{array}{c} \sin\theta\\ \pm1 \\ \cos\theta \end{array}\right),
\label{eq5}
\end{equation}
where $\tan\theta=v/w$. For convenience, and in analogy to the quantum case, we will refer to these eigenvalues and eigenstates as \textit{instantaneous} eigenenergies and eigenstates in the following.

\begin{figure}
\centering
\includegraphics[width=0.235\textwidth]{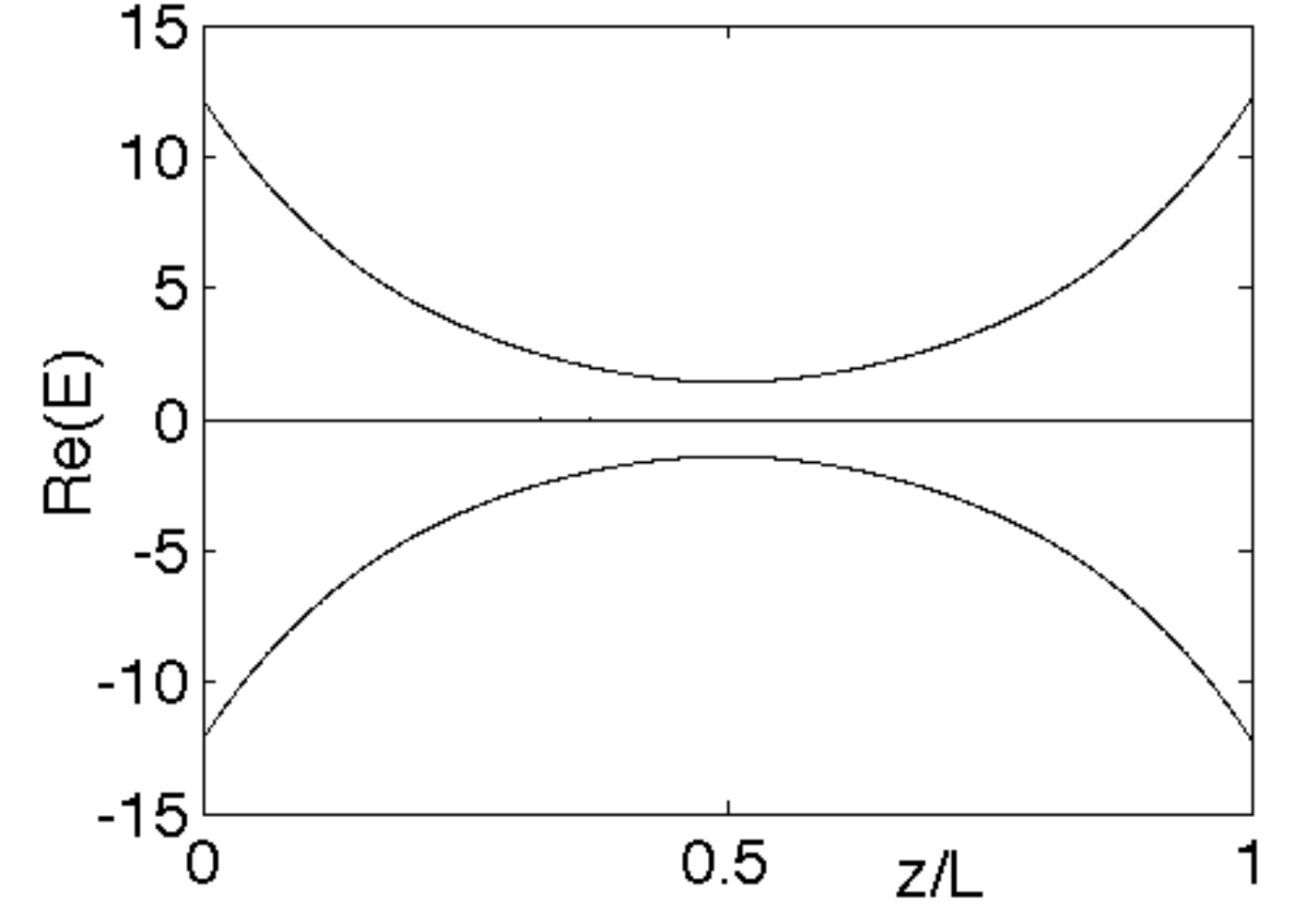}
\includegraphics[width=0.235\textwidth]{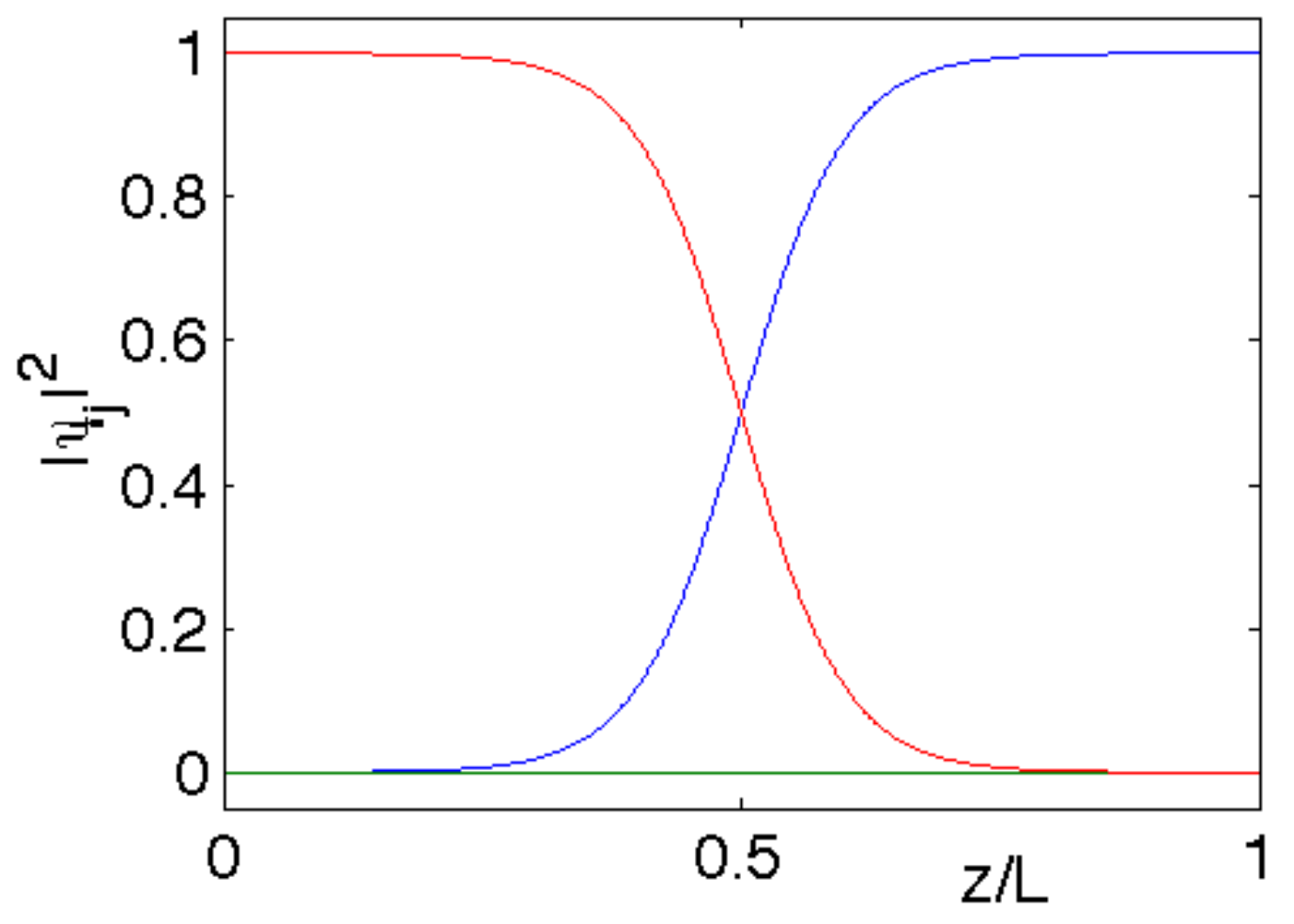}
\caption{The left figure shows the eigenvalues of the Hamiltonian (\ref{eq_Ham}), with the parameter dependence (\ref{eq2}) for $a=5$. The right figure shows the squared absolute values of the components of the adiabatic eigenstate $|\varphi^0\rangle$ (first, second, and third component shown in blue, green, and red, respectively).}
\label{fig1}
\end{figure}

An example of the instantaneous eigenvalues of this system is depicted in the left panel in Fig.~\ref{fig1}. The right panel shows the components of the instantaneous eigenstate $|\varphi^0\rangle$ corresponding to the zero energy eigenvalue. As in the conventional STIRAP scheme, $|\varphi^0\rangle$ varies from $|3\rangle$ (only populating the right waveguide) at $z=0$ to $|1\rangle$ (only populating the left waveguide) at $z=L$ for sufficiently large values of $a$, for which 
\begin{equation}
\frac{w(0)}{v(0)}=\frac{v(L)}{w(L)}=e^{-a}\ll1.
\end{equation} 
Thus, adiabatic parameter variation, which is given for sufficiently large values of $L$ in the present case, leads to a complete population transfer from the right to the left waveguide. Furthermore, since the second component of $|\varphi^0\rangle$ is identically zero for all $z$, the central waveguide is never significantly populated during the process. This has been experimentally demonstrated in the waveguide setup in \cite{Lahi08}. 

\subsection{Adiabatic transfer probability}
\label{subsec_trans_prob}
Let us now consider the transfer probability in more detail. 
We wish to adiabatically follow the state $|\varphi^0\rangle$, which corresponds approximately to the right waveguide at $z=0$, that is,
\begin{equation}
|\psi(0)\rangle = |\varphi^0(0)\rangle\approx -|3\rangle.
\end{equation}
The transfer probability to the left waveguide at $z=L$ is defined by
\begin{equation}
\label{eqn_transprob_def}
P=\frac{|\langle 1| \psi(L)\rangle|^2}{{\sum_{n=1}^3 |\langle n|\psi(L)\rangle|^2}},
\end{equation}
where the denominator is constant if there is no absorption, and can be set to unity.

Let us express the solution of the evolution equation (\ref{eqn_evo}) as a sum of adiabatic and non-adiabatic parts in the instantaneous eigenbasis:
\begin{equation}
|\psi(z)\rangle = a_{0}(z) |\varphi^0(z)\rangle+\sum_{j=\pm} a_{j}(z)|\varphi^{j}(z)\rangle,
\label{eq14}
\end{equation}
where the absolute values of the nonadiabatic coefficients of the instantaneous states $|\varphi^{\pm}\rangle$ are equal due to the symmetry of the problem. On account of the relation $|\varphi^0(L)\rangle\approx|1\rangle$, the transfer probability in Eq. (\ref{eqn_transprob_def}) can be estimated as
\begin{equation}
\label{eqn_transprob}
P=\frac{|a_{0}(L)|^2}{|a_{0}(L)|^2+2|a_{\pm}(L)|^2}.
\end{equation}
Again, if there is no absorption in the system, the denominator is equal to unity if the initial wave function is normalised. 

The dynamical equations for the coefficients $a_j$ directly follow from (\ref{eqn_evo}) as
\begin{equation}
\frac{da_{j}}{dz} 
= -iE_ja_j
-\sum_{k \ne j}
\langle \bar{\varphi}^{j}|\frac{d\varphi^{k}}{dz}\rangle
a_{k}
\label{new4}
\end{equation}
with the initial conditions
\begin{equation}
a_0(0) = 1,\quad a_\pm(0) = 0.
\label{new5}
\end{equation}
The bar denotes complex conjugation 
distinguishing left and right eigenvectors for 
the symmetric possibly non-Hermitian Hamiltonian to be considered in the following. In that case the geometric phase is included in the eigenvectors by using the normalisation condition $\langle\bar \varphi^j|\varphi^j\rangle=1$ \cite{Mois11}.
If there is no absorption the Hamiltonian under consideration is real symmetric, and the instantaneous eigenstates can be chosen real; thus no complex conjugation is required. For the Hamiltonian (\ref{eq_Ham}) the equations (\ref{new4}) explicitly read
\begin{equation}
i\frac{d}{dz}\!\left(\!\begin{array}{c}a_-\\
a_0\\
 a_+\end{array}\!\right)\!
=\!\left(\!\begin{array}{ccc}
-\omega & i\frac{\sqrt{2}a}{L\,\omega^2} & 0\\
- i\frac{\sqrt{2}a}{L\,\omega^2}   & 0 & - i\frac{\sqrt{2}a}{L\,\omega^2} \\
0 & i\frac{\sqrt{2}a}{L\,\omega^2}   &\omega 
 \end{array}\!\right) \!\!
 \left(\!\begin{array}{c}a_-\\
a_0\\
 a_+\end{array}\!\right),
\label{new10a}
\end{equation}
with $\omega=\sqrt{v^2+w^2}$, which depends on $z$. For large values of L the non-adiabatic coupling constant $\frac{\sqrt{2}a}{L\,\omega^2}$ is small for all values of $z$, and reaches its maximal value at  $z=L/2$. 

\begin{figure}
\centering
\includegraphics[width=0.49\textwidth]{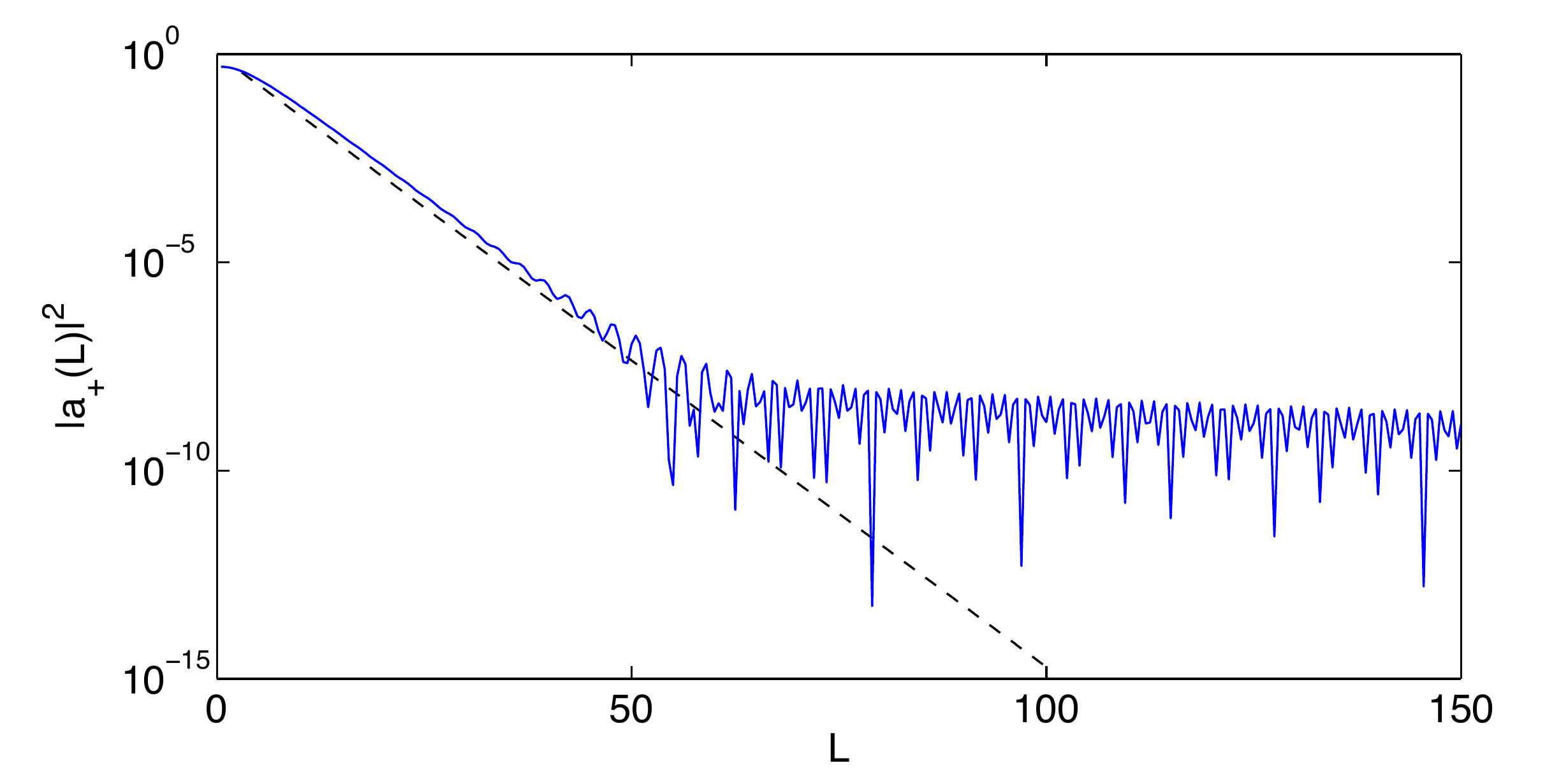}
\caption{Nonadiabatic transition probabilities as a function of $L$ for $a=5$ (blue solid line). For comparison also the estimation based on equation (\ref{eq10}) is shown (black dashed line).}
\label{fig2}
\end{figure}

The nonadiabatic transition probability can be obtained by a numerical integration. The result is depicted in a semilogarithmic plot as a function of $L$ in Fig. \ref{fig2}, for $a=5$. While there is an approximate exponential decay initially, at a critical value of $L$ the transition probability begins to oscillate around an only slowly decreasing mean value. Similar behaviour is typically observed in STIRAP schemes \cite{Lain96}. 

The initial exponential decay of the non-adiabatic transition probability in Fig. \ref{fig2} can be estimated by a Landau-Zener type argument as
\begin{equation}
P_{nonad} \approx \exp\left(-2\mathrm{Im}\,\int_{L/2}^{z_0} (E_{+}-E_{0}) dz\right),
\label{eq8}
\end{equation}
where $z_0$ denotes the position of the exceptional point, where $E_0=E_+$ or $E_0=E_-$, nearest to the real $z$-axis \cite{Landau}. Since $E_-=E_+$, exceptional points appear as triple degeneracies (EP3) with $E_+ = E_0 = E_- = 0$, corresponding to complex values of $z$ given by the equation $v = \pm iw$. Using (\ref{eq2}), we find that they are located at
\begin{equation}
z_n = L\left(\frac{1}{2}+i(1+2n)\frac{\pi}{4a}\right), \quad 
n \in \mathbb{Z}. 
\label{eq6}
\end{equation}
The EP3 with the smallest positive imaginary part of $z$ is $z_0$. From $E_0 = 0$ and 
$E_+(L/2+i\xi) = \sqrt{v^2+w^2} =  \sqrt{2\cos (2a\xi/L)}$, we can compute the integral in (\ref{eq8}) as
\begin{eqnarray}
\nonumber P_{nonad} &\approx& \exp\left(-2\int_{0}^{\mathrm{Im}\,z_0} 
\sqrt{2\cos (2a\xi/L)} d\xi\right) \\
&=&\exp\left(-\frac{2}{a\sqrt{\pi}}\Gamma^2\!\left({\textstyle\frac{3}{4}}\right)L\right).
\label{eq10}
\end{eqnarray}
This estimate is shown in Fig. \ref{fig2} as a dashed black line. It can be seen that it provides a good approximation for the initial decrease of the transition probability with increasing $L$. The saturation of the transition probability with larger values of $L$ is due to the fact that the Landau-Zener approximation assumes a vanishing coupling at the beginning and the end of the ``time evolution'' (i.e. at $z=0$ and $z=L$, respectively), which is not the case in the STIRAP scheme investigated here. 

In summary, the transfer probability (\ref{eqn_transprob}) is very close to unity as long as $L$ is large. In the following we will see how the scheme can break down in the presence of losses, even though the non-adiabatic transitions stay small. 

\section{Breakdown of adiabaticity in the presence of absorption}
The adiabatic theorem states that a system initially prepared in an eigenstate remains in the corresponding instantaneous eigenstate if the system parameters are varied infinitely slowly and the corresponding energy level  is nondegenerate at all times. For finite parameter variations, there are small transition amplitudes between the adiabatic states, which are typically of order $O(\epsilon)$, where $\epsilon$ denotes the small adiabatic parameter \cite{Berr90,Hage02}. We have seen for the STIRAP scheme in waveguides investigated here, that indeed the non-adiabatic transitions are negligible for large values of $L$. In the presence of absorption, however, this alone does not guarantee adiabatic evolution. This is due to the fact, that absorption leads to complex eigenvalues, that is, the amplitudes of the instantaneous eigenstates are themselves exponentially decreasing in time. This can lead to a situation in which the small nonadiabatic transition amplitude grows exponentially in time relative to the adiabatic amplitude, if the adiabatic state is not the one with the smallest decay rate. In other words, the effect is caused by the dominance of the single gain mode of the time evolution operator \cite{nhermad}. We will now demonstrate how this general phenomenon can lead to a breakdown of the adiabatic population transfer in our waveguide system, if one of the waveguides has an additional absorption. Note that this does not rule out the existence of a modified transfer scheme which could lead to a total population transfer in the presence of absorption. See, e.g., \cite{Kuhn} for investigations of such alternative schemes in the context of cavity QED.

The case of absorption in the central waveguide is well studied in the context of the original STIRAP scheme, and it is straight forward to show, that it does not influence the transition probability significantly. Intuitively this can be understood by realizing that the adiabatic state $|\varphi^0\rangle$ is not affected by the additional decay, and thus, does not populate the central waveguide at any instant. For moderate absorption the non-adiabatic coupling elements are not altered and thus the system will simply not feel the absorption in the unpopulated state in a first order approximation. In the following we shall focus on the effect of absorption in one of the outer waveguides, which can indeed lead to a breakdown of the transfer property.

\begin{figure}
\centering
\includegraphics[width=0.235\textwidth]{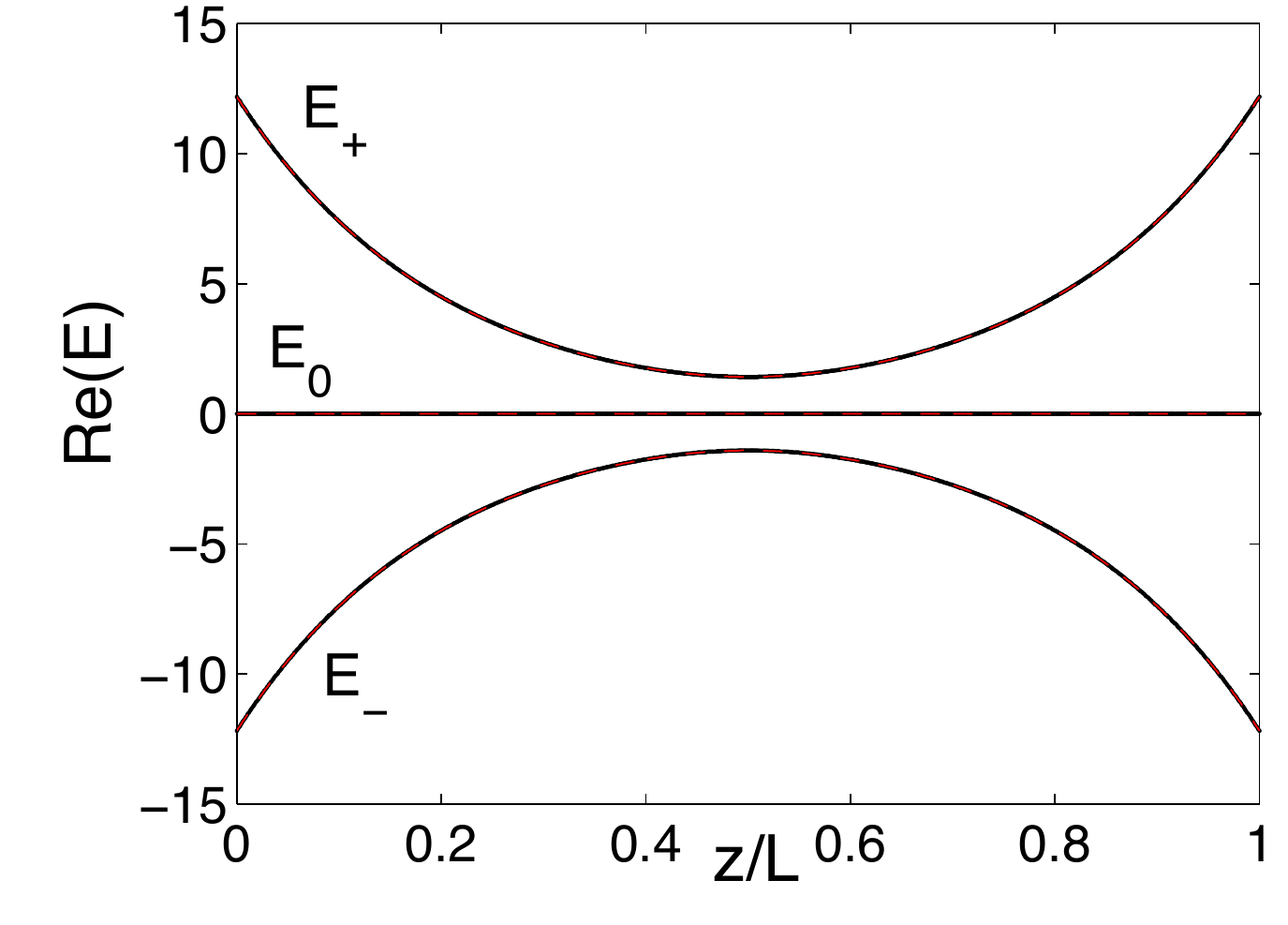}
\includegraphics[width=0.235\textwidth]{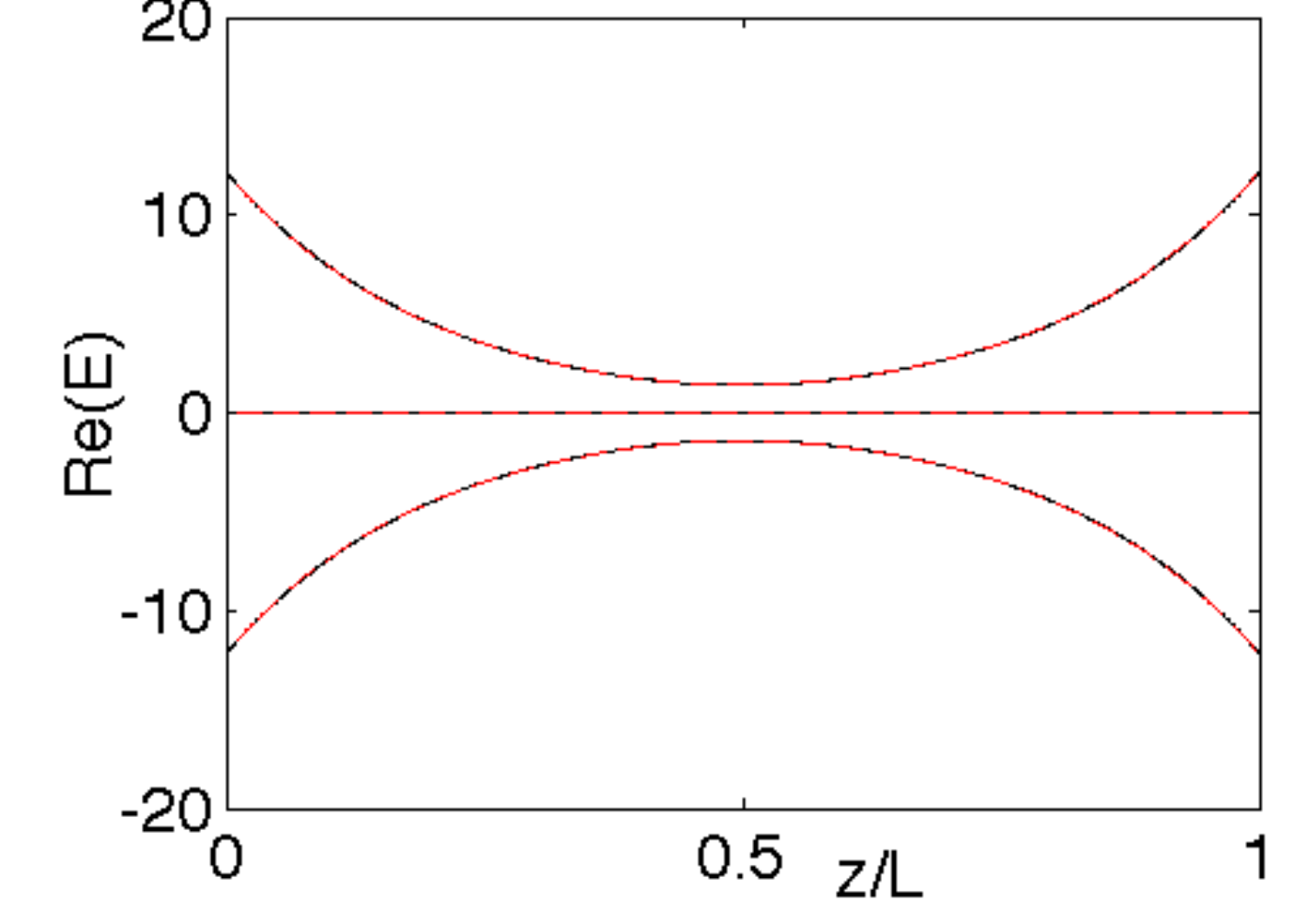}
\includegraphics[width=0.235\textwidth]{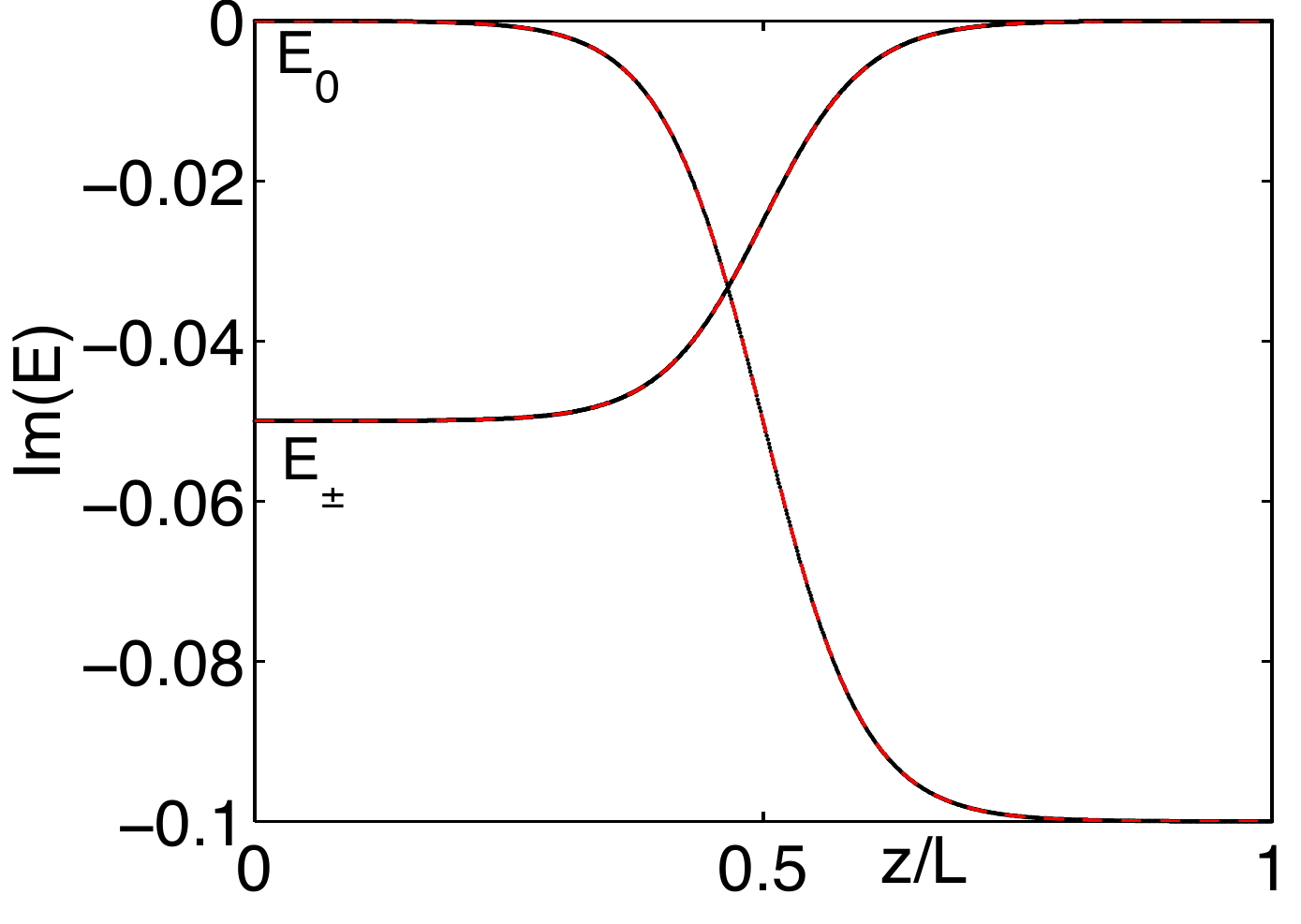}
\includegraphics[width=0.235\textwidth]{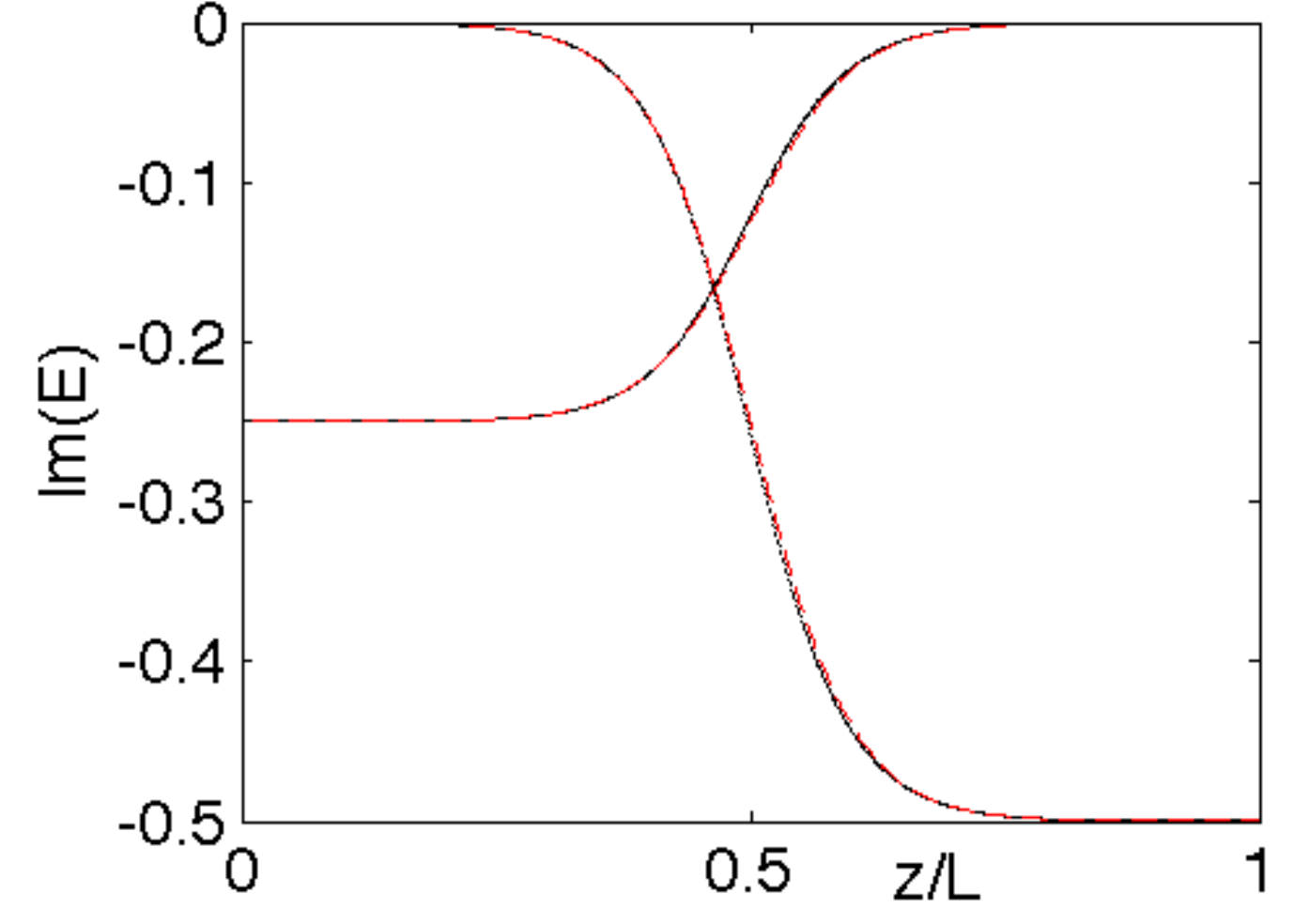}
\includegraphics[width=0.235\textwidth]{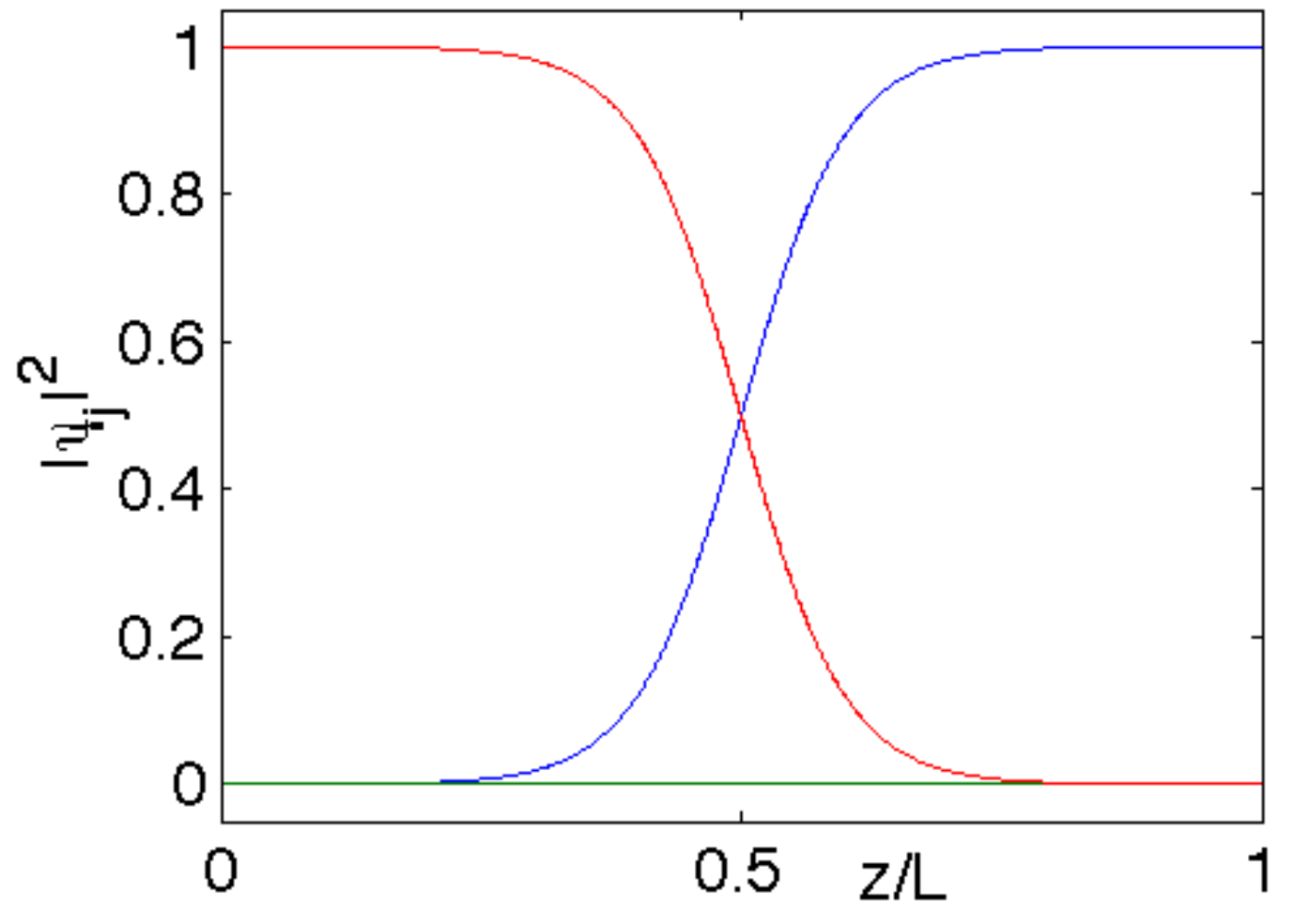}
\includegraphics[width=0.235\textwidth]{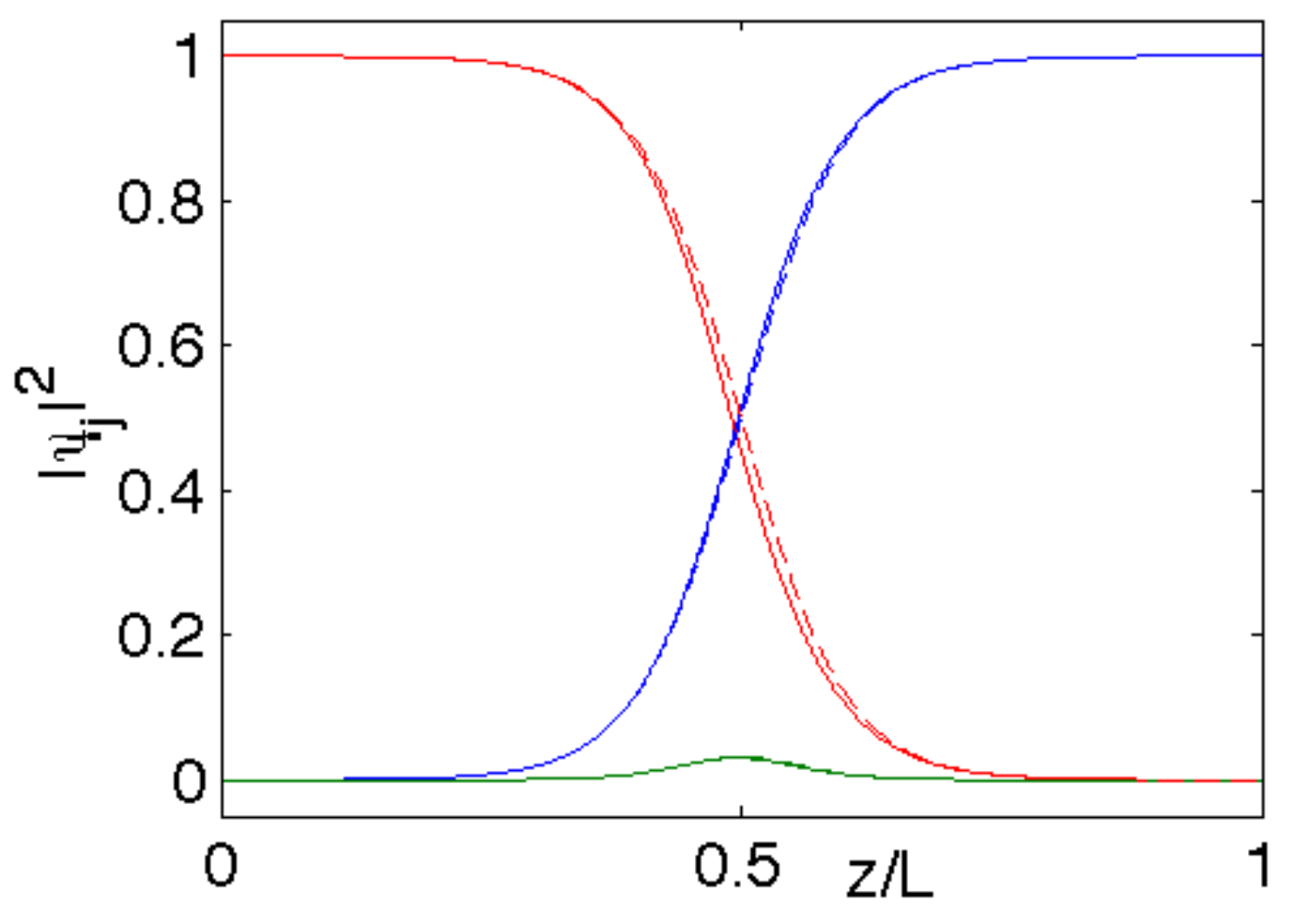}
\caption{Real (top) and imaginary (middle) parts of the energies for $\gamma=0.1$ (left) and $\gamma=0.5$ (right), and $a=5$. The solid black line shows the numerically obtained eigenvalues, and the dashed red line those obtained from first order perturbation theory. The solid lines in the lower panel show the components of the adiabatic eigenstate $|\varphi^0\rangle$ (colours as in figure 1), the dashed lines show the perturbative result.}
\label{fig3}
\end{figure}

\subsection{Approximate eigenvalues and eigenstates in the presence of absorption}
Absorption in the $n$-th waveguide is modelled by an imaginary energy $-i\gamma$ in the $n$-th diagonal element of the Hamiltonian (\ref{eq_Ham}). Here we are in particular interested in the effect of small absorption rates, which only slightly modify the eigenvalues and eigenstates. In this case, the eigenvalues can be obtained via first order perturbation theory that leaves the real parts unaltered and yields additional imaginary parts \cite{Mois11,Seyr03}. For absorption in the target waveguide (state $|1\rangle$), i.e. for the Hamiltonian 
\begin{equation}
H(z) = \left(\begin{array}{ccc}
-i\gamma & \,v(z) & 0 \\
\,\,\,v(z) & 0 & w(z) \\
\,\,\,0 & \,w(z) & 0
\end{array}\right),
\label{eq_Ham2}
\end{equation} 
we find:
\begin{eqnarray}
\nonumber{\rm Im}(E_0)&\approx&-\gamma\cos^2\theta=-\gamma\frac{w^2}{\omega^2},\\
\label{eqn-eig-pert}{\rm Im}(E_{\pm})&\approx &-\gamma\frac{1}{2}\sin^2\theta=-\gamma\frac{v^2}{2\omega^2}.
\end{eqnarray}
The eigenstate corresponding to $E_0$, in first order perturbation theory, is given by
\begin{equation}
|\varphi^0\rangle = \left(\begin{array}{c} \cos\theta \\ i\gamma\cos\theta\sin\theta/\sqrt{v^2+w^2} \\ -\sin\theta \end{array}\right).
\end{equation}
In the upper and middle panels of Fig.~\ref{fig3} we show the numerically obtained energy levels in comparison to the first order perturbation theory for $a=5$, and two different values of $\gamma$ as a function of $z/L$. The energies are well described by the perturbative equation. In the lower panels of the figure the components of the instantaneous eigenstate $|\varphi^0\rangle$ are shown. As in the Hermitian case, this state populates the right waveguide for $z=0$, while it is localized in the left for $z=L$. However, the central waveguide is also partly populated for intermediate values of $z$. 

A similar behaviour is found for absorption in the initial waveguide (state $|3\rangle$), described by the Hamiltonian 
\begin{equation}
H(z) = \left(\begin{array}{ccc}
0 & \,v(z) & 0 \\
\,\,\,v(z) & 0 & w(z) \\
\,\,\,0 & \,w(z) & -i\gamma
\end{array}\right),
\label{eq_Ham3}
\end{equation} 
where the eigenvalues acquire an additional imaginary part of 
\begin{equation}
{\rm Im}(E_0) \approx -\gamma\sin^2\theta,\quad
{\rm Im}(E_\pm) \approx-\frac{\gamma}{2}\cos^2\theta.
\label{eqn-eig-pert2}
\end{equation}

Assuming that the non-adiabatic transitions are small, one could conclude from the behaviour of the eigenvalues and eigenstates that starting in the right waveguide ($|3\rangle$) for large values of $L$ a full transfer of the population to the left waveguide ($|1\rangle$) (up to an overall decay) should be achieved for moderate values of $\gamma$. However, a numerical simulation of the total transferred probability, as a function of $\gamma$ and $L$, yields a different result, as depicted in Fig. \ref{fig3} for absorption in the target waveguide and in Fig. \ref{fig6} for absorption in the intitial waveguide: In both cases there is a sharp transition for relatively small values of $\gamma$ at which the adiabatic transfer scheme breaks down. This is related to the relative growth of the nonadiabatic transitions mentioned above. In the case of absorption in the initial waveguide, there are additional separate regions where the population is transferred completely for larger values of $\gamma$. These, however, correspond to non-adiabtic (coherent) transfer. In what follows we shall first investigate in detail how the adiabatic transition boundary emerges for absorption in the target waveguide, before turning to the case of absorption in the initial waveguide in section \ref{subsec_initial_decay}.

\begin{figure}
\centering
\includegraphics[width=0.45\textwidth]{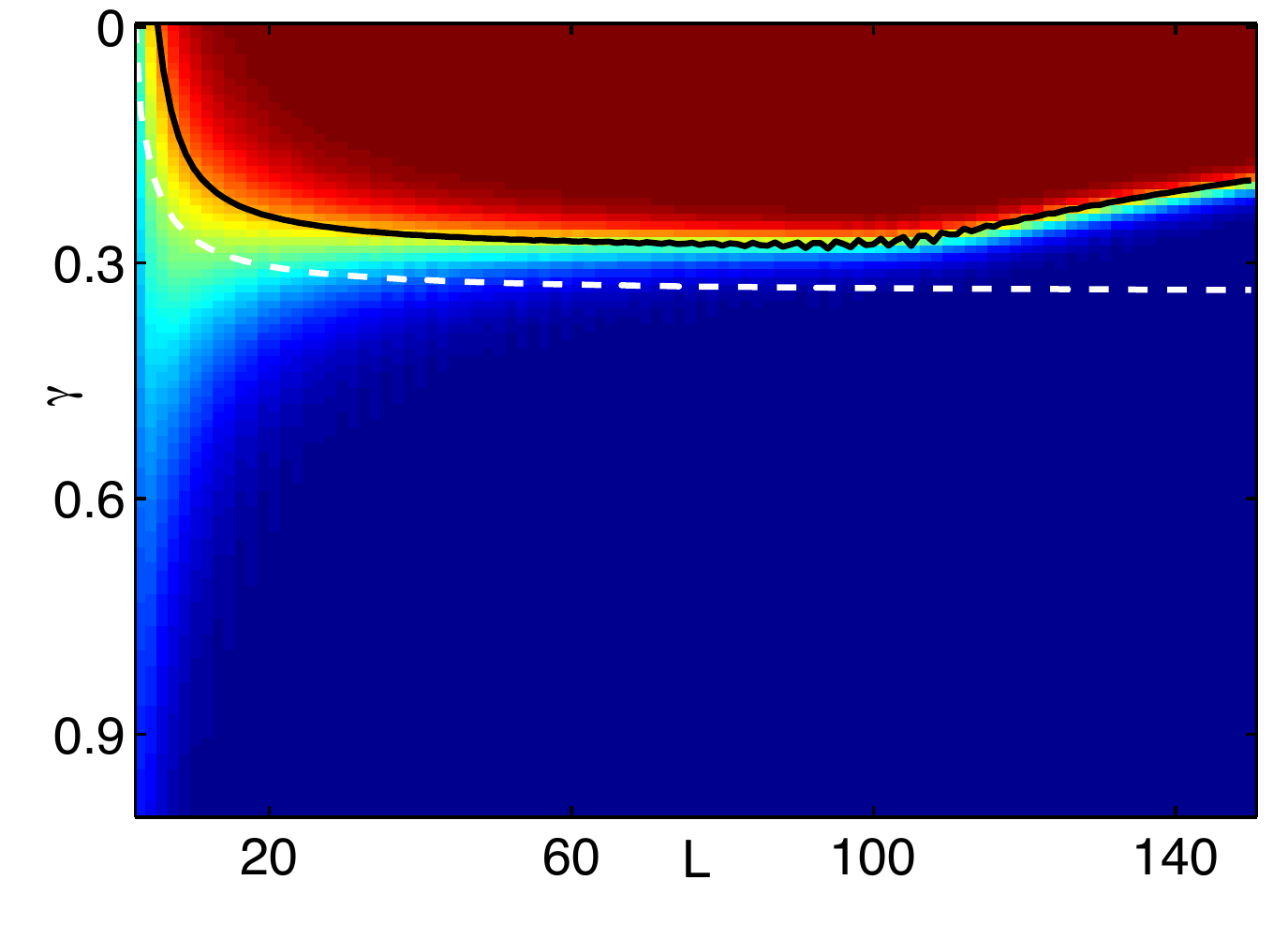}
\includegraphics[width=0.45\textwidth]{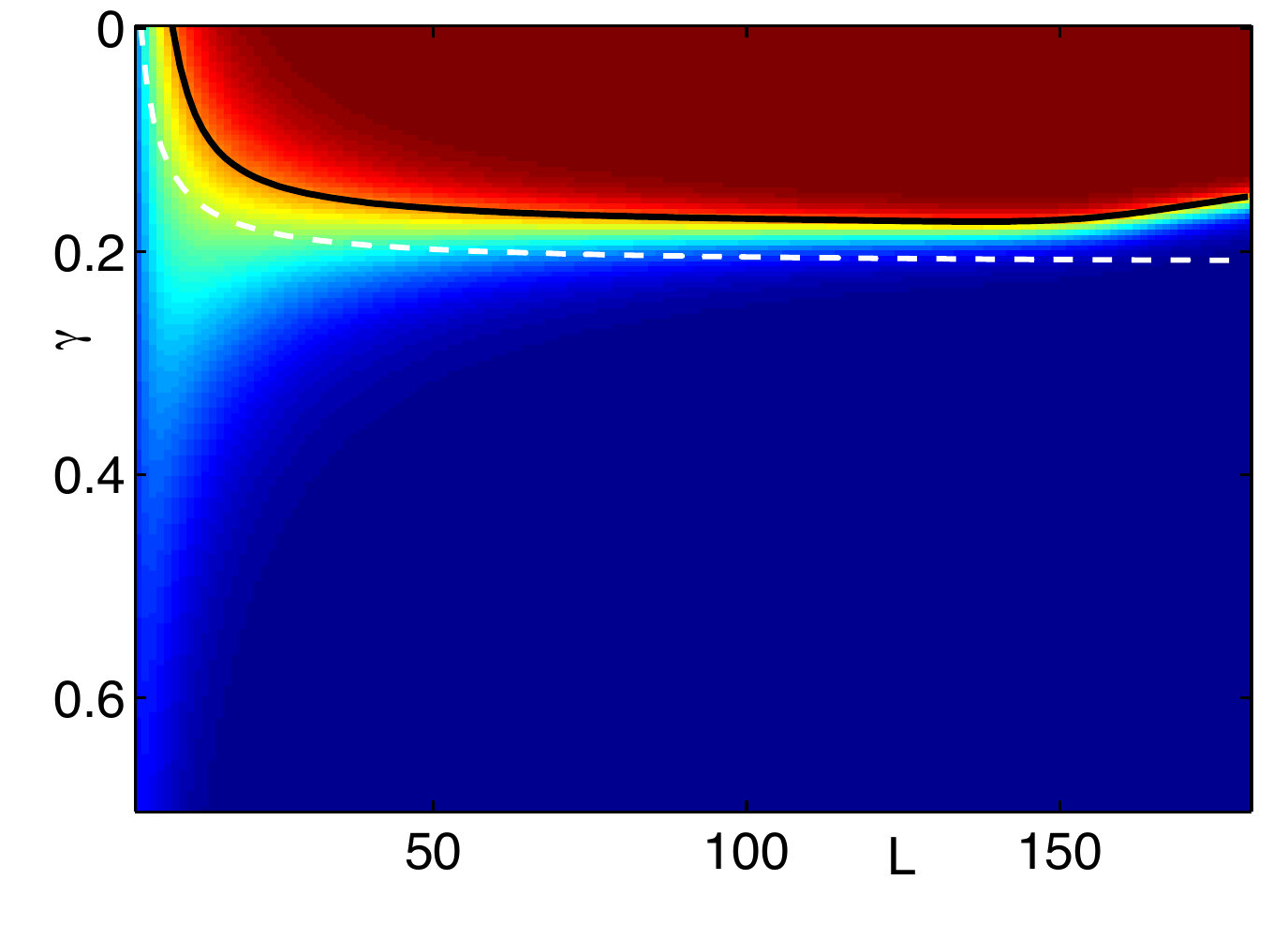}
\caption{Numerically obtained transfer probability ($P = 1$ is red, $P = 0$ is blue) for absorption in the target waveguide as a function of $L$ and $\gamma$ for $a=5$ (top) and $a=8$ (bottom) . The dashed white line shows the analytical transition boundary based on the amplification of the nonadiabatic coupling for the Hermitian system using the approximation (\ref{eq10}), the solid black line is the semianalytical critical boundary using the numerically obtained nonadiabatic coupling for the non-Hermitian system with $\gamma= 0.2$.}
\label{fig4}
\end{figure}

\subsection{Absorption in the target waveguide}
As before we express the time-dependent state in the basis of the instantaneous eigenvectors according to equation (\ref{eq14}), and analyse the behaviour of the adiabatic coefficients. The main difference to the absorption free case discussed in section \ref{subsec_trans_prob} is that  we have to take into account that the adiabatic states have a finite lifetime, that is, they decay.  Approximating the instantaneous eigenvectors by those of the Hermitian system, while at the same time, using the first order corrections for the 
energies (\ref{eqn-eig-pert}) to account for the decay rates,
equations (\ref{new4}) become
\begin{equation}
i\frac{d}{dz}\!\left(\!\begin{array}{c}a_-\\
a_0\\
 a_+\end{array}\!\right)\!
=\!\left(\!\begin{array}{ccc}
-\omega -i\frac{\gamma\, v^2}{2\omega^2} & i\frac{\sqrt{2}a}{L\,\omega^2} & 0\\
- i\frac{\sqrt{2}a}{L\,\omega^2}   & -i\frac{\gamma\, w^2}{\omega^2} & - i\frac{\sqrt{2}a}{L\,\omega^2} \\
0 & i\frac{\sqrt{2}a}{L\,\omega^2}   &\omega  -i\frac{\gamma\, v^2}{2\omega^2} 
 \end{array}\!\right) \!\!
 \left(\!\begin{array}{c}a_-\\
a_0\\
 a_+\end{array}\!\right).
\label{eq_nonad_coupl1}
\end{equation}
The non-adiabatic coupling constant is not affected by the absorption, and still is maximal in the vicinity of the avoided crossing of the energies, close to $z=L/2$ (see Fig. \ref{fig3}). The adiabatic state $|\varphi^0\rangle$ is approximately stable for $z<L/2$ where ${\rm Im}(E_0)$ is small, and decays for $z>L/2$, where ${\rm Im}(E_0)\approx-\gamma$. Hence, we can assume that the main transition to the non-adiabatic states $|\varphi^{\pm}\rangle$ still occurs in the neighborhood of $z\approx L/2$. After this point, the states $|\varphi^{\pm}\rangle$ decay only slowly since the imaginary parts of their energies are small. 
Thus we find
\begin{equation}
|a_{\pm}(L)| 
\approx \sqrt{P_{nonad}},
\label{eq11}
\end{equation}
where $P_{nonad}$ denotes the nonadiabatic transition probability. 
The remaining population in the state $|\varphi^0\rangle$, which we wish to follow adiabatically, on the other hand, decays after the transition. Thus we estimate 
\begin{eqnarray}
\nonumber \left|a_{0}(L)\right| 
&=&\sqrt{1-2P_{nonad}}\exp\left(\int_{0}^L \mathrm{Im}\,E_0dz\right) \\
&\approx& \sqrt{1-2P_{nonad}}\,{\rm e}^{-\gamma L/2},
\label{eq12}
\end{eqnarray}
where the integral is calculated using the eigenvalues in first order perturbation theory in equation (\ref{eqn-eig-pert}). The factor of $2$ in front of the nonadiabatic transition probability accounts for transitions into the two states $|\varphi^\pm\rangle$.

The population transfer is successful if 
\begin{equation}
|a_{0}(L)|^2 \gg 2|a_{\pm}(L)|^2,
\label{eq23}
\end{equation}
that is, when the non-adiabatic transition probability is small.
Treating (\ref{eq23}) as an equality and using (\ref{eq11}) and (\ref{eq12}), we obtain the threshold value 
\begin{equation}
\label{eqn_gamma_cr}
\gamma_{cr}=\ln{\left(\frac{1}{2P_{nonad}}-1\,\right)}/L.
\end{equation}
On account of the exponential dependence of $a_0$ on $\gamma$ in Eq. (\ref{eq12}), the characteristic width of the threshold can be estimated by $\delta\gamma_{cr} \sim 2/L$.
Since $\delta\gamma_{cr} \ll \gamma_{cr}$ for small $P_{nonad}$ in Eq. (\ref{eqn_gamma_cr}),
the observed threshold appears to be rather sharp.

Since we expect $\gamma_{cr}$ to be relatively small, to get a first estimate we approximate the nonadiabatic transition probability with its initial exponential behaviour in the Hermitian case where $\gamma=0$, given by equation (\ref{eq10}). Substituting this expression into (\ref{eqn_gamma_cr}) yields an analytic estimate for the critical decay rate,
\begin{equation}
\gamma_{cr}^{LZ}= \frac{1}{L} \ln \left( \frac{ \exp{\left(
\frac{2}{a\sqrt{\pi}} \Gamma^2\left( \frac{3}{4}\right)L\right)}}{2} - 1\right) ,
\end{equation}
which for large values of $L$ simplifies to
\begin{equation}
\gamma_{cr}^{LZ}\approx\frac{2}{a\sqrt{\pi}}\Gamma^2\!\left({\textstyle\frac{3}{4}}\right)-\frac{\ln(2)}{L}.
\end{equation}
In Fig. \ref{fig4} this value is shown as a white dashed line. It can be seen that it is a good estimate for the exact transition boundary. It fails, however, to describe a slow decrease of $\gamma_{cr}$ for large values of $L$.
This decrease is due to the non-exponential behaviour of the transition probability for large values of $L$, already observed in Fig. \ref{fig2}.

\begin{figure}
\centering
\includegraphics[width=0.49\textwidth]{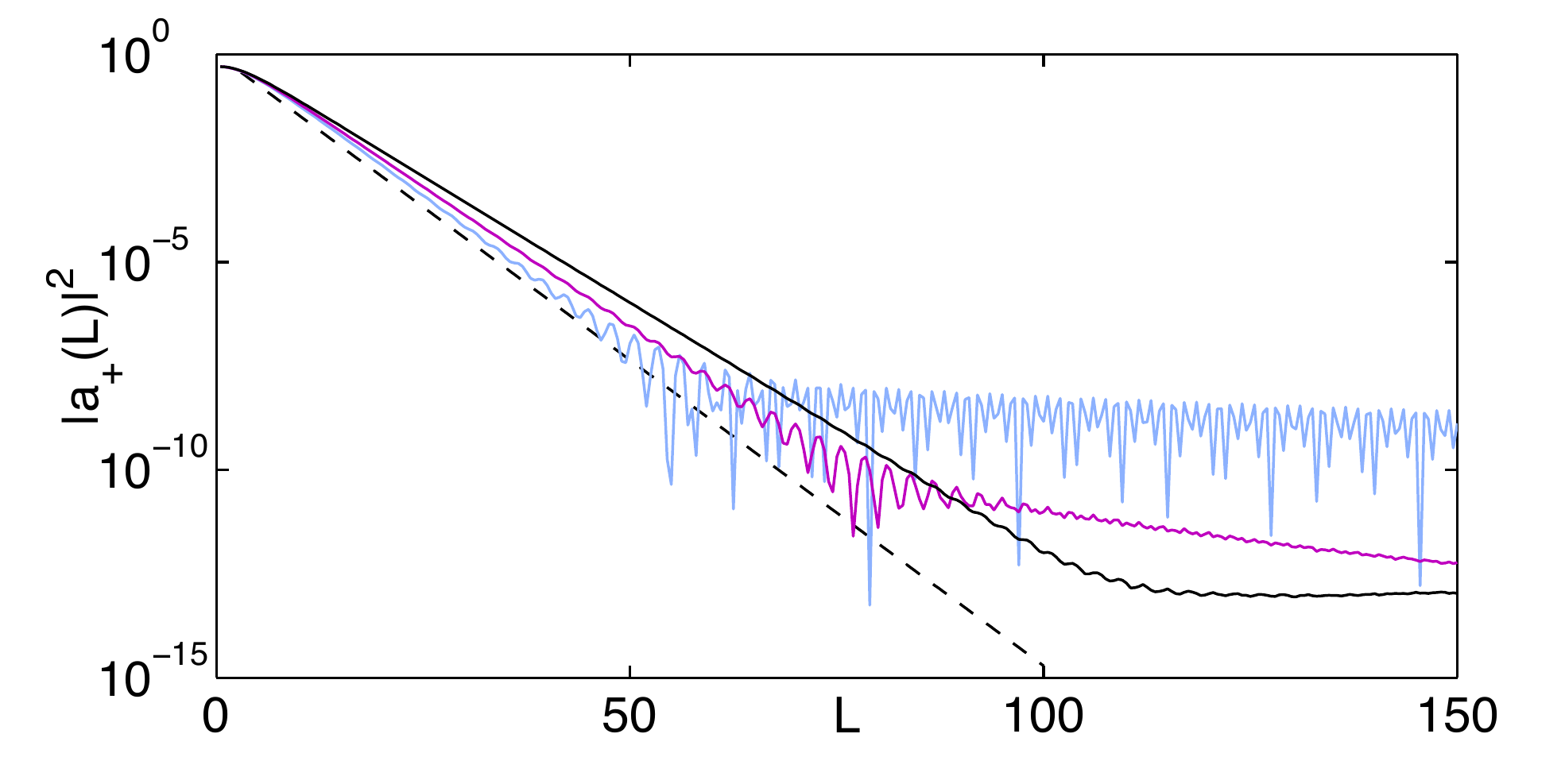}
\caption{Nonadiabatic transition probabilities as a function of $L$ for $a=5$ for different values of $\gamma$: $\gamma=0$ (light blue), $\gamma=0.1$ (magenta) and $\gamma=0.25$ (black). For comparison also the estimation based on equation (\ref{eq10}) is shown (black dashed line).}
\label{fig5}
\end{figure}

The exact nonadiabatic transition probability can again be obtained by a numerical integration. We show the result for three different values of $\gamma$ in a semilogarithmic plot as a function of $L$ in Fig. \ref{fig5}. The light blue line corresponds to the Hermitian case. For comparison, the Landau-Zener estimate (\ref{eq10}) is also shown (dashed black line). The presence of absorption leads to a smoothing of the oscillations, and an increase of the value of $L$ for which the initially approximately exponential decay starts to saturate. It can also be seen that the slope in the region of exponential decay is slightly decreased by the absorption. The latter effect explains why the white lines in Fig. \ref{fig4} slightly overestimate the critical value of $\gamma$. As expected the quality of the approximation is better for the case $a=8$, where the boundary is located at smaller values of $\gamma$. 

We can obtain an excellent approximation for the transfer boundary by using the numerical values of the nonadiabatic transition probability $P_{nonad}$ for a fixed value of $\gamma=0.2$, close to the actual boundary, as an input in equation (\ref{eqn_gamma_cr}). This is demonstrated in Fig. \ref{fig4}, where the solid black lines depict the thus obtained result. 

\subsection{Absorption in the intial waveguide}
\label{subsec_initial_decay}
Let us finally analyse the case with decay in the initial state, modelled by 
the Hamiltonian (\ref{eq_Ham3}).
In this case, similar to the case of absorption in the target waveguide, using the decay rates in Eq. (\ref{eqn-eig-pert2}), we find the dynamical equations for the coefficients of the state in the instantaneous 
eigenbasis:
\begin{equation}
i\frac{d}{dz}\!\left(\!\begin{array}{c}a_-\\
a_0\\
 a_+\end{array}\!\right)\!
=\!\left(\!\begin{array}{ccc}
-\omega -i\frac{\gamma\, w^2}{2\omega^2} & i\frac{\sqrt{2}a}{L\,\omega^2} & 0\\
- i\frac{\sqrt{2}a}{L\,\omega^2}   & -i\frac{\gamma\, v^2}{\omega^2} & - i\frac{\sqrt{2}a}{L\,\omega^2} \\
0 & i\frac{\sqrt{2}a}{L\,\omega^2}   &\omega  -i\frac{\gamma\, w^2}{2\omega^2} 
 \end{array}\!\right) \!\!
 \left(\!\begin{array}{c}a_-\\
a_0\\
 a_+\end{array}\!\right).
\label{eq_nonad_coupl2}
\end{equation}

While superficially these dynamical equations appear very similar to (\ref{eq_nonad_coupl1}), in contrast to the previously considered cases, the only important nonadiabatic coupling occurs for small $z/L$. The non-adiabatic coupling constant $\frac{\sqrt{2}a}{L\,\omega^2}$ is the same in both cases, and still has a maximum around $z=L/2$. However, the adiabatic state $|\varphi^0\rangle$ is now exponentially decaying for values of $z<L/2$, and the actual transition is proportional to the population in the adiabatic state. Thus, the loss of population due to absorption can cancel out the increase in the coupling constant for increasing values of $z$. This is the case for large values of $L$ and sufficiently large absorption $\gamma\gg1/L$.

To quantitatively estimate for which values of $\gamma$ the adiabatic transfer breaks down, as before we have to estimate at which value of $\gamma$ the condition $|a_{0}(L)|^2 \gg 2|a_{\pm}(L)|^2$ holds. 

\begin{figure}
\centering
\includegraphics[width = 0.49\textwidth]{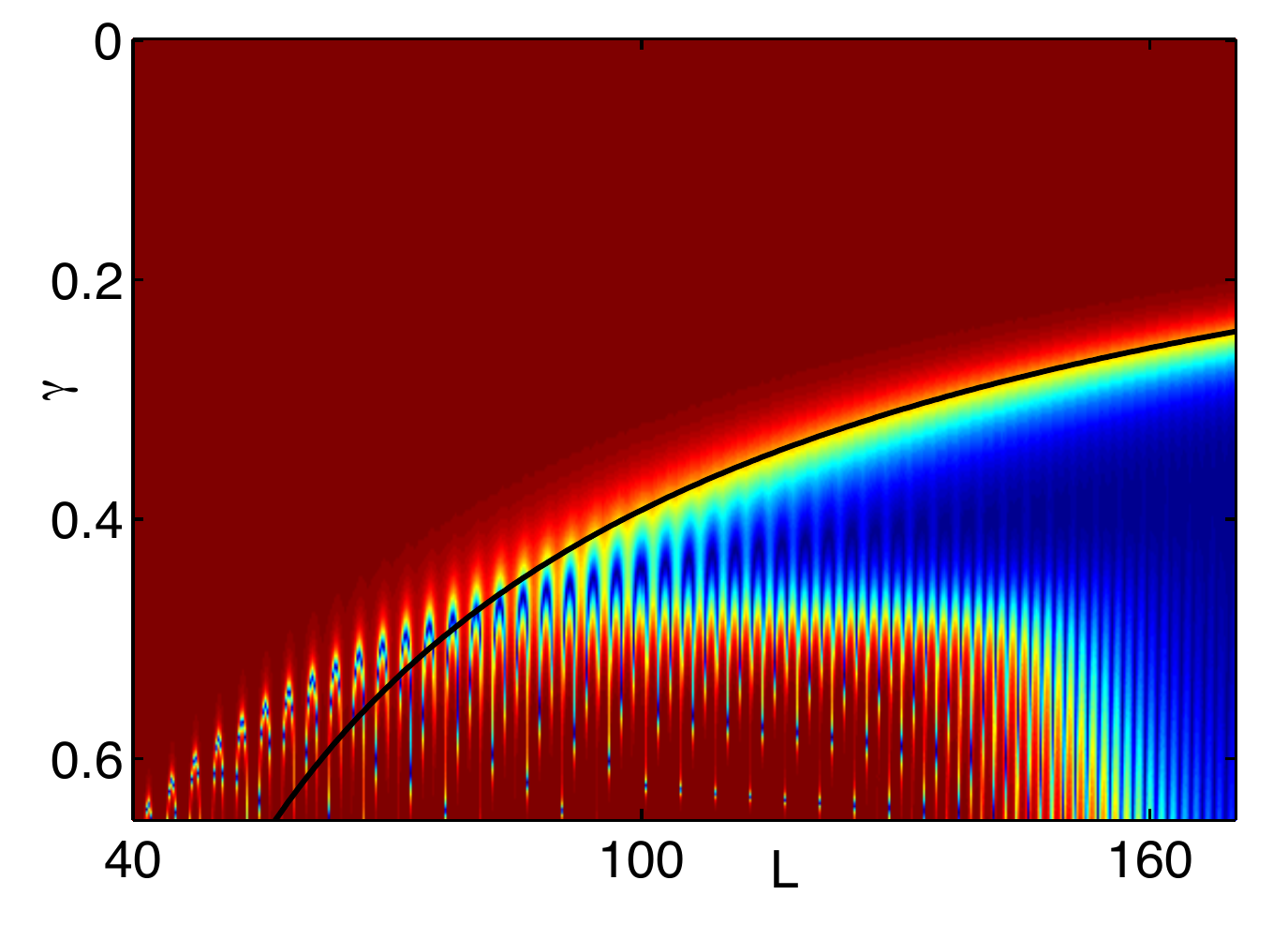}
\caption{Numerically obtained transfer probability ($P = 1$ is red, $P = 0$ is blue) for absorption in the initial waveguide as a function of $L$ and $\gamma$ for $a=5$. The black line shows the analytical transition boundary (\ref{new16}).}
\label{fig6}
\end{figure}

For small values of $z\ll L$, we can use Eq. (\ref{eq2}) and set
\begin{equation}
\frac{v^2}{\omega^2} \approx 1, \quad
\frac{w^2}{\omega^2}\approx e^{-2a} \ll 1, \quad
\omega \approx e^{a/2}\gg 1
\label{new20}
\end{equation}
in equations (\ref{eq_nonad_coupl2}). If we further neglect a small nonadiabatic effect for the amplitude $a_0$, which is justified by the initial conditions (\ref{new5}), the dynamical equations (\ref{eq_nonad_coupl2}) reduce to
\begin{equation}
\frac{da_0}{dz} 
= 
-\gamma a_0,\quad
\frac{da_\pm}{dz} 
= 
\mp i e^{a/2} a_\pm
+\frac{ae^{-a}\sqrt{2}}{L} a_0.
\label{new11}
\end{equation}
Using (\ref{new5}) this system is solved by
\begin{equation}
a_0 = e^{-\gamma z},
\end{equation}
and
\begin{equation}
a_\pm = 
-\frac{ae^{-a}\sqrt{2}}{(\pm i e^{a/2}-\gamma)L}e^{\mp i e^{a/2}z}
\left[1-e^{(\pm i e^{a/2}-\gamma) z}\right].
\label{new12}
\end{equation}
And thus
\begin{equation}
|a_\pm| = \frac{ae^{-a}\sqrt{2}}{\sqrt{\gamma^2+e^a}L}\sqrt{1-2\cos(e^{a/2}z)e^{-\gamma z}+e^{-2\gamma z}}.
\label{new13a}
\end{equation}
The term $\sqrt{1-2\cos(e^{a/2}z)e^{-\gamma z}+e^{-2\gamma z}}$ oscillates around the value $1+e^{-2\gamma z}$ with decreasing amplitude. For values of $z\gg1/\gamma$, $|a_{\pm}|$ thus converges to the constant value 
\begin{equation}
|a_\pm| = \frac{ae^{-a}\sqrt{2}}{\sqrt{\gamma^2+e^a}L}.
\label{new13}
\end{equation}
Since the population in the adiabatic state continues to decrease exponentially up to $z\sim L/2$ and is consequently small afterwards, the nonadiabatic transitions for larger $z$ are small 
and do not change the final result at $z = L$ for sufficiently large $L\gg1/\gamma$.  
The only important process for values of $z>L/2$ is the decay with the rates given in (\ref{eqn-eig-pert2}). 
Using (\ref{new13}), this yields
\begin{equation}
|a_0(L)| = e^{\int_0^L \mathrm{Im}\,E_0 dz} = e^{-\gamma L/2},
\label{new14}
\end{equation}
and
\begin{eqnarray}
\nonumber |a_\pm(L)| &=& 
\frac{ae^{-a}\sqrt{2}}{\sqrt{\gamma^2+e^a}L}
e^{\int_0^L \mathrm{Im}\,E_\pm dz} \\
&=&\frac{ae^{-a}\sqrt{2}}{\sqrt{\gamma^2+e^a}L}e^{-\gamma L/4}.
\label{new14b}
\end{eqnarray}
The adiabatic transfer breaks down when $|a_0|^2 \approx 2|a_{\pm}|^2$, which yields the condition
\begin{equation}
e^{-\gamma_{cr} L} \approx
\frac{4a^2e^{-2a}}{(\gamma_{cr}^2+e^a)L^2}e^{-\gamma_{cr} L/2}.
\label{new15}
\end{equation}
Since $e^a\gg1$, while we expect the critical value of $\gamma$ to be small, we can neglect the term $\gamma_{cr}^2$ in the denominator in Eq. (\ref{new15}), to obtain an approximation for 
the critical value of $\gamma$ at which adiabatic transfer breaks down:
\begin{equation}
\gamma_{cr} =\frac{4}{L}\log
\frac{L}{2ae^{-3a/2}}.
\label{new16}
\end{equation}
Again, due to the exponential dependence on $\gamma_{cr}$ in Eq. (\ref{new15}) the breakdown of adiabatic transfer has a relatively sharp threshold of width $\delta\gamma_{cr}\sim2/L\ll\gamma_{cr}$. Figure~\ref{fig6} shows the boundary given by (\ref{new16}) together with the results of direct numerical simulation, confirming the validity of our derivation. The deviation for smaller values of $L$ is due to the fact, that the assumption of the only relevant non-adiabatic transitions occurring for small values of $z/L$ is only justified for sufficiently large values of $L$. 

\section{Conclusion} We have demonstrated that even a small decay rate can significantly influence the dynamical behaviour of a system with respect to adiabatic time evolutions. This is due to a competition between small nonadiabatic transition amplitudes and relative exponential growths of the decaying adiabatic eigenstates. In particular, we have shown for a STIRAP related scheme, which can be implemented straightforwardly using optical waveguides, that the adiabatic transfer behaviour breaks down at a relatively sharp threshold for small decay rates. The critical value of the decay rate has been estimated by simple analytical arguments.

\vskip 9pt 
\noindent The authors thank Raam Uzdin for stimulating discussions and helpful comments. EMG and AAM are grateful to the Lewiner Institute for Theoretical Physics and the Department of Chemistry at the Technion, Israel for support and hospitality during this work. EMG acknowledges support via the Imperial College JRF scheme. AAM acknowledges support from CNPq (Grant Nr 305519/2012-3). NM  acknowledges support from ISF (Grant Nr 298/11).

\end{document}